# Dynamics of Hadronization from Nuclear Semi Inclusive Deep Inelastic Scattering

Kawtar Hafidi – on behalf of the CLAS collaboration

Physics Division, Argonne National Laboratory, Argonne, IL 60439, USA

The CLAS experiment E02-104, part of the EG2 run at Jefferson Lab, was performed to study the hadronization process using semi inclusive deep inelastic scattering off nuclei. Electron beam energy of 5 GeV and the CLAS large acceptance detector were used to study charged pion production. The high luminosity available at Jefferson Lab and the CLAS large acceptance are key factors for such measurements allowing high statistics and therefore multidimensional analyses of the data. Both the multiplicity ratio and the transverse momentum broadening for carbon, iron and lead relative to deuterium are measured. Preliminary results for positive pions are discussed.

## Introduction

Hadronization is the mechanism by which the quark, when ejected from a nucleon neutralizes its color to from many hadrons in the final state. The description of hadronization in terms of fragmentation functions is known in the vacuum, but the physical information on its dynamics is still a mystery. The time-space evolution of the hadron formation is unknown due to the fact that it is a non perturbatif process and therefore is hard to calculate at the most fundamental level. Unfortunately, Lattice QCD calculations are not yet helpful in this matter. To access this dynamic, the nucleus will play the role of the detector since the production length is comparable with the nuclear size. The nuclear environment provides a unique opportunity to look at the early stage of hadronization a few Fermi from the origin. Therefore, by studying the properties of leading particles emerging from deep inelastic scattering (DIS) on a range of nuclei, important information on the characteristic time distance-scales of hadronization can be determined as a function of several variables.

## The in-medium hadronization

The hadronization process in the nuclear medium is traditionally described in the framework of phenomenological string models[1] and final-state interactions of the produced hadron with the surrounding medium. Several models with different views exist on the market, which all introduce various time scales. While some of them emphasize the interaction of the struck quark with the partons in the medium[2], claiming the hadron is produced outside the nucleus, others describe hadronization in terms of both parton and hadron interactions[3]. Experimental measurements are needed to pin down the time scale characteristics of the color neutralization process, the nature of the intermediate state necessary to the hadron formation, and the mechanism of their interaction with the nuclear medium.

## Experimental tools

A primary experimental observable is the hadronic multiplicity ratio:

$$R_M^h(z, \nu, p_T^2, Q^2) = \left\{ \frac{N_h^{DIS}(z, \nu, p_T^2, Q^2)}{N_e^{DIS}(\nu, Q^2)} \right\}_A \Bigg/ \left\{ \frac{N_h^{DIS}(z, \nu, p_T^2, Q^2)}{N_e^{DIS}(\nu, Q^2)} \right\}_D$$

$N_h^{DIS}$ and $N_e^{DIS}$ denote the number of hadrons and electrons measured in DIS kinematics, $\nu$ is the virtual photon energy, $z = E_h/\nu$ is the fraction of the photon energy carried by the hadron, $Q^2$ is the four-momentum transfer, and $p_T$ is the transverse momentum of the hadron relative to the virtual photon direction. The second experimental quantity is called the transverse momentum broadening $\langle \Delta p_T^2 \rangle = \langle p_T^2 \rangle_A - \langle p_T^2 \rangle_D$, where $\langle p_T^2 \rangle$ is the $p_T^2$ averaged. The dependence of the multiplicity ratio on several kinematical variables offers a wealth of information on quark propagation in the nuclear medium and hadron formation. Studying the transition from high to low $\nu$ dependence is correlated with the transition from quark propagation to hadron propagation in nuclei while the z dependence controls the broadening effects. In addition, the transverse momentum broadening provides valuable information on the parton energy loss which is believed to play an important role in predicting signatures of the quark gluon plasma at RHIC[4] through the phenomenon of jet quenching[2].

## Measurements and preliminary results

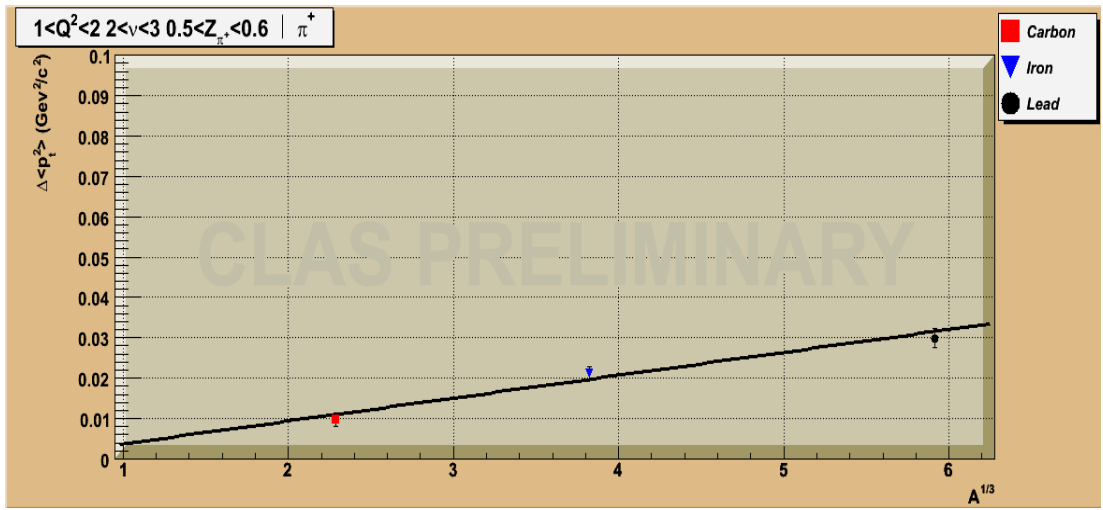

**Figure 1:** Preliminary $\pi^+$ data from CLAS EG2 run for $\Delta p_T^2$ vs. $A^{1/3}$ for carbon, iron and lead for one bin in $Q^2$, z, and $\nu$. The line is a linear fit.

Jefferson Lab offers unique capabilities to study hadron production in DIS. Measurements with the CEBAF Large Acceptance Spectrometer (CLAS) provide data on the widest possible range of nuclear target masses at high luminosity. Data at 5 GeV were taken in 2003 with $10^{34}$ cm$^{-2}$s$^{-1}$ luminosity using two targets simultaneously on (the liquid deuterium target and one solid target; alternating between C, Fe and Pb) to reduce systematic errors in the multiplicity ratio. Figure 1 shows preliminary $\pi^+$ data from the CLAS EG2 experiment at 5.0 GeV. These data are for z from 0.5 to 0.6, $Q^2$ from 1 to 2 GeV$^2$ and $\nu$ from 2 to 3 GeV. The plot shows a linear behavior of the transverse momentum broadening as a function of the nuclear radius within the statistical uncertainties. These are the first direct and precise measurements of $p_T$ broadening on several nuclei, confirming that the quark energy loss has a quadratic dependence on the quark path length in the nuclear medium in this kinematical regime. This is consistent with the non Abelian analog of the LPM effect in QED[5].

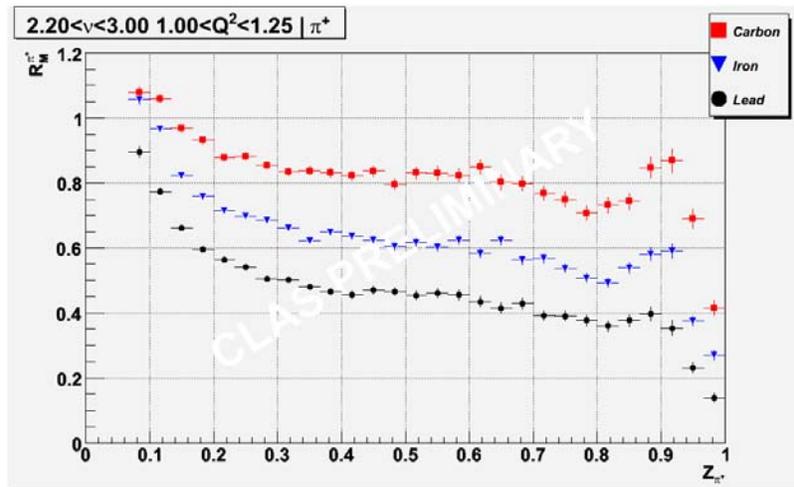

**Figure 2:** Preliminary data from the CLAS EG2 run for the hadronic multiplicity ratio $R_M^{\pi^+}$ for positive pions in carbon (top), iron, and lead (bottom).

An example of preliminary hadronic multiplicity ratio results is given in Figure 2. The plot is for $Q^2$ from 1 to 1.25 GeV$^2$ and $\nu$ from 2.2 to 3 GeV for C, Fe and Pb. The high statistics collected in this experiment allow measuring the multiplicity ratio with significantly smaller uncertainties than HERMES experiment[6].

## Conclusions

Recently, the EG2 experiment at CLAS has taken data at 5 GeV and obtained an unprecedented statistics for hadronization from carbon, iron, and lead. Preliminary results for positive pions have been briefly discussed. Analyses of kaons and negative pions are underway. The presented data show strong attenuation effects, and follow similar trends as the HERMES data. The precision of EG2 data will further constrain theoretical models. The planned JLab 12 GeV upgrade and the CLAS12 detector will offer an order of magnitude more luminosity ($10^{35}$ $cm^{-2}s^{-1}$) and improved particle identification capabilities. These improvements will permit mapping out multivariate $R_M^h$ for many hadrons ($\pi$, K, $\eta$, $\phi$, p, $\Lambda$, $\Xi$ and $\Sigma$) as well as $p_T$ broadening for many of these, spanning a kinematical range of $\nu$ from 2 – 9 GeV and 2 – 9 $GeV^2$ in $Q^2$.

## Acknowledgements

I would like to acknowledge the great efforts by Hayk Hakobyan, the graduate student who performed the analysis. I would like also to acknowledge the outstanding efforts of the staff of the Accelerator and the Physics Divisions and the technical staff is Hall B at Jefferson Lab that made this experiment possible. This work was supported by the Italien Instituto Nazionale di Fisica Nucleare, The Frensh Centre National de la Recherche scientifique, the French Commissariat a l'Energie Atomique, the U.S. Department of Energy and National Science Foundation, the UK Engineering and Physical Science research Council, and the Korean Research Foundation. The Southeastern Universities Research Association (SURA) operates the Thomas Jefferson National Accelerator Facility for the United States Department of Energy under contract DE-AC05-84ER40150.